\documentclass[final,3p,times,twocolumn]{elsarticle}

\usepackage{amssymb}
\usepackage{amsmath,amssymb,exscale}
\usepackage{graphicx}
\usepackage{epsfig}
\usepackage{multicol}
\usepackage{color}
\usepackage{xcolor}
\usepackage{hyperref}
\usepackage{mathrsfs}
\usepackage{hyperref}
\hypersetup{colorlinks,bookmarksopen,bookmarksnumbered,citecolor=rossos,
linkcolor=black,pdfstartview=FitH,urlcolor=blus}
\usepackage{amsfonts}
\usepackage{slashed}
\usepackage{blindtext}
 \usepackage{fancyhdr}
\usepackage{hyperref}
\usepackage{mathtools}
\biboptions{numbers,sort&compress}

\allowdisplaybreaks

\definecolor{blus}{cmyk}{1,1,0,0.}
\definecolor{verdes}{cmyk}{0.99,0,0.99,0.02}
\definecolor{rossos}{cmyk}{0,1,1,0.55}
\definecolor{greeny}{cmyk}{0.99,0,0.59,0.98}
\definecolor{redy}{cmyk}{0,1,1,0.40}

\newcommand{\bp}{\bar M_{\rm Pl}}

\def\be{\begin{equation}}
\def\ee{\end{equation}}
\def\bea{\begin{eqnarray}}
\def\eea{\end{eqnarray}}
\def\ba{\begin{array} }
\def\ea{\end{array}}
\def\bac{\begin{array} {c}}
\def\bacc{\begin{array} {cc}}
\def\baccc{\begin{array} {ccc}}

\def\psl{\hbox{\hbox{${p}$}}\kern-1.9mm{\hbox{${/}$}}}
\def\dsl{\hbox{\hbox{${\partial}$}}\kern-1.7mm{\hbox{${/}$}}}
\def\Dsl{\hbox{\hbox{${D}$}}\kern-2.1mm{\hbox{${/}$}}}
\def\intg{\int \hspace{-.1cm}d^4\hspace{-.05cm}x\hspace{-.05cm}\sqrt{-g}}
\definecolor{red}{rgb}{1,0,0}

\def\hhref#1{\href{http://arxiv.org/abs/#1}{arXiv:#1}}  
\hypersetup{colorlinks=true, linkcolor=red}

\journal{the arXiv}

\begin{document}

\begin{frontmatter}

  %to do
  %resubmit to PRD (checking "references" is present)
 
\title{\vspace{-2cm}\huge {\color{redy}Metastability in Quadratic Gravity}} 
%Chaotic Initial Conditions and Quadratic Gravity
%Chaotic Inflation and Quadratic Gravity
%Chaotic pre-inflationary initial conditions and Quadratic Gravity
%
%Ghost metastability in Quadratic Gravity

\author{\vspace{1cm}{\large {\bf Alberto Salvio}}}

\address{\normalsize \vspace{0.2cm}Physics Department, University of Rome and INFN Tor Vergata, Italy \\

\vspace{0.3cm}
% {\it {\small Report numbers: }}
 \vspace{-1cm}
 }

\begin{abstract}
Quadratic gravity is a UV completion of general relativity, which also solves the hierarchy problem.
 The presence of 4 derivatives implies via the Ostrogradsky theorem  that the {\it classical} Hamiltonian is unbounded from below.
 Here we solve this issue by showing that the relevant solutions are not unstable but metastable. When the energies are much below a threshold (that is high enough to describe the whole cosmology) runaways are avoided. Remarkably, the chaotic inflation theory of initial conditions ensures that such bound is satisfied and we work out  testable implications for the early universe. 
  The possible instability occurring when the bound is violated  not only is compatible with cosmology but would also explain why we live in a homogeneous and isotropic universe.
\end{abstract}

% \end{keyword}

\end{frontmatter}
 
 \begingroup
\hypersetup{linkcolor=blus}
\tableofcontents
\endgroup

%\newpage

\section{Introduction}\label{introduction}
Let us start with some basic definitions. The  action of quadratic gravity (QG) in the Jordan frame is (modulo total derivatives)
\be S=\intg\left(\frac{R^2}{6f_0^2}-\frac{W^2}{2f_2^2} - \frac{\bp^2R}{2}+ \mathscr{L}_m \right),\label{S}   \ee
where $W^2\equiv W_{\mu\nu\rho\sigma}W^{\mu\nu\rho\sigma}$, $W_{\mu\nu\rho\sigma}$ is the Weyl tensor and $\mathscr{L}_m$ is the matter piece
 (a cosmological constant can be included, but we neglect it here given its tiny value). $\mathscr{L}_m$ includes non-minimal couplings between the scalar fields $\phi^a$ and the Ricci scalar of the form $-\xi_{ab}\phi^a\phi^b R/2$. The parameters $f_0^2$ and $f_2^2$ are positive to avoid tachyons\footnote{In this work the flat metric $\eta_{\mu\nu}$ has the mostly minus signature: $\eta_{\mu\nu} =$ diag$(+1,-1,-1,-1)$.}~\cite{Stelle:1976gc,Salvio:2018crh}. 
 
 The $R^2$ and $W^2$ terms render gravity renormalizable~\cite{Stelle:1976gc}. However, renormalizability requires the space of states to be endowed with an indefinite inner product. It was recently realized that this does not preclude a physical interpretation as the probabilities involving observable states can  be computed with positive norms~\cite{Salvio:2015gsi,Salvio:2018crh,Strumia:2017dvt} (see also~\cite{Hawking:2001yt,Bender:2007wu,Bender:2008gh,Holdom:2016xfn,Salvio:2016vxi,Raidal:2016wop,Anselmi:2017yux,Donoghue:2017fvm,Anselmi:2017ygm,Salvio:2018kwh,Anselmi:2018bra,Anselmi:2019rxg, Donoghue:2018izj} for related approaches).
 
 Another remarkable feature of QG is the possibility to solve the hierarchy problem (why is the Higgs mass much smaller than $\bp$?): this requires\footnote{One could go up to $f_2\sim10^{-7}$ with specific matter contents, but we quote here the most general bound.}  $f_2\lesssim 10^{-8}$~\cite{Salvio:2014soa,Giudice:2014tma,Kannike:2015apa,Salvio:2017qkx}. Note that, as will be reviewed in the article, the Starobinsky inflationary model~\cite{Starobinsky:1980te} (which is built in here thanks to the presence of the $R^2$ term) requires a very small $f_0$ to match the observed curvature power spectrum; this is another independent reason to think that a very small value of $f_2$ is natural.

Still the classical theory may harbour some issues: given that QG features 4 time-derivatives in the Lagrangian, Ostrogradsky theorem~\cite{ostro} establishes that the classical
%\footnote{Renormalizability ensures that all quantum fluctuations carry positive energies and, therefore, the vacuum state is stable~\cite{Stelle:1976gc,Salvio:2015gsi,Salvio:2018crh}.}
 Hamiltonian is not bounded from below. This manifests itself through the presence of a ghost with spin 2 and mass~\cite{Stelle:1976gc,Stelle:1977ry}
 \be  M_2=\frac{f_2\bp}{\sqrt{2}},\label{M2def}\ee
 which is due to the $W^2$ term.  In the present  paper this problem will be addressed. 
 
 The basic idea  exploits two key elements: ({\it i}) the ghost is not tachyonic, ({\it ii}) its coupling $f_2$~\cite{Salvio:2017qkx}  is very small. Indeed, as well-known, a non-tachyonic decoupled ghost does not suffer from any instability~\cite{Pais,Salvio:2015gsi}. By introducing an order one coupling to normal particles one expects, from effective field theory arguments, that the theory remains stable up to  energies of order $M_2$ (below which the ghost is not excited). But, given that the ghost coupling is tiny to solve the hierarchy problem, this energy threshold is lifted to a much higher value. In the rest of the paper we confirm this expectation. 
 
 At the end, we also work out testable predictions for inflation.

  \vspace{-0.3cm}

\section{Ghost metastability}\label{model}

\subsection{A 4-derivative scalar field example}\label{A 4-derivative scalar field example}
In order to illustrate our argument in a clear way we start by presenting it in a 4-derivative theory of a real scalar field $\phi$. Indeed the potential issues due to the Ostrogradsky theorem are present in this simple case too. In Sec.~\ref{The case of quadratic gravity} we will then turn to QG.

 The Lagrangian is given by
\be  \mathscr{L}_\phi= - \frac{1}{2}\phi \Box \phi - \frac{c_4}{2} \phi \Box^2 \phi - V(\phi)\label{Lag-simple},\ee
where $c_4$ is a real parameter, $\Box\equiv \eta^{\mu\nu}\partial_{\mu}\partial_\nu$  and the function $V$   represents some interaction. At low energy the second term in~(\ref{Lag-simple}) is negligible and one obtains the standard Lagrangian of a scalar field. 

The 4-derivative terms  in (\ref{Lag-simple}) can be eliminated by introducing an $A$uxiliary field $A$: one adds 
\be \frac{c_4}{2}\left(\Box \phi -\frac{A-\phi/2}{c_4}\right)^2\ee
 (that is zero by using the field equation of $A$) to $ \mathscr{L}_\phi$, which then becomes
\bea \hspace{-2.4cm}\mathscr{L}_\phi \hspace{-0.2cm}&=& \hspace{-0.3cm}- \frac{1}{2}\phi \Box \phi - \frac{c_4}{2} \phi \Box^2 \phi+\frac{c_4}{2}\left(\Box \phi -\frac{A-\phi/2}{c_4}\right)^2 - V(\phi)
 \nonumber\\
&=&-A \Box\phi+ \frac1{2c_4}\left(A-\frac{\phi}2\right)^2 - V(\phi). \label{Lstep}\eea
One can diagonalize the kinetic terms by defining $\phi_\pm\equiv \phi/2\pm  A$,  that is 
\be A\equiv \frac{1}{2}(\phi_+-\phi_-)\qquad \phi\equiv \phi_+ + \phi_-,\ee
%phi is the sum of the coordinate, not of one coordinate and one momentum like in the transformation from the initial and tilded variables in the 4 derivative oscillator. However, one can bring it in that form through a canonical transformation.
to obtain
\be\mathscr{L}_\phi=-\frac12 \phi_+\Box\phi_++\frac12 \phi_-\Box\phi_- + \frac{m^2}{2}\phi_-^2 - V(\phi_+ + \phi_-),  \label{Lagsimple2} \ee
where $m^2\equiv 1/c_4$. The corresponding  field equations are
  \be \Box \phi_+= - V'(\phi_++\phi_-), \qquad \Box \phi_-= - m^2\phi_-^2 + V'(\phi_++\phi_-). \ee
We observe that the theory includes two 2-derivative scalars, which  are decoupled in the non-interacting case $V=0$: a standard one $\phi_+$, which is massless and a ghost $\phi_-$ with mass $m$ (note that $c_4>0$ in order for $\phi_-$ not to be a tachyon, a condition which we assume here). $\phi_+$ and $\phi_-$ are analogous to the massless graviton and the ghost in QG, respectively.

  \begin{figure}[t]
\begin{center}
  \includegraphics[scale=0.5]{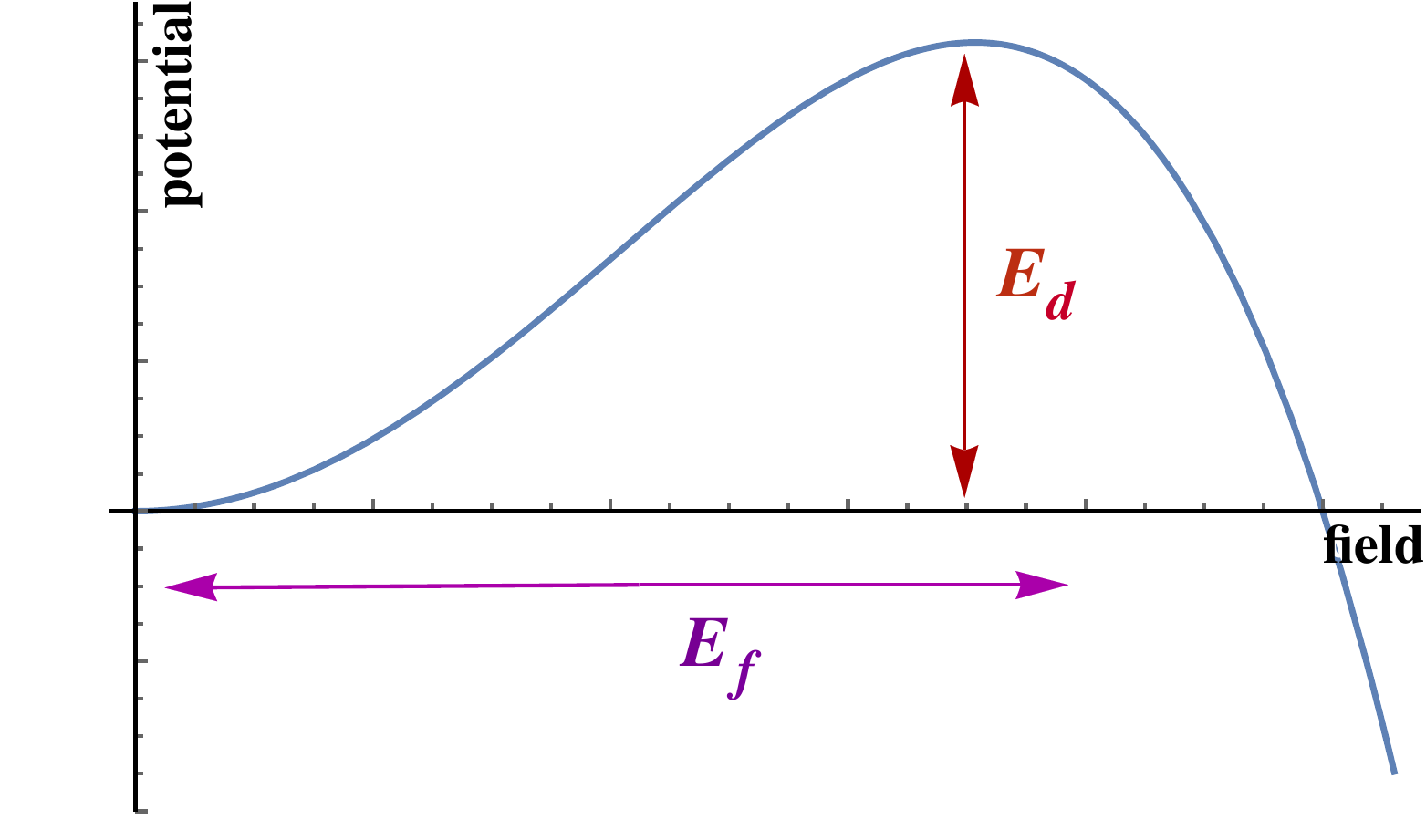} 
  \end{center}
   \caption{\em   Geometrical meaning of the thresholds $E_f$ and  $E_d$  (defined in~(\ref{Efdef}) and~(\ref{Eddef}))  for field values and derivatives,  respectively.}
\label{drawPot}
\end{figure}

   Let us now assume some non-trivial interaction: $V(\phi) = \lambda \phi^4/4$, where $\lambda$ is a positive coupling constant. Then we see that  $V$ tends to stabilize the motion of $\phi_+$ and destabilize the one of $\phi_-$. If $m^2\leq 0$  the solution $\phi_+ = \phi_- =0$ would be unstable. However, for $m^2> 0$ the situation is much better as long as the values and the derivatives of the fields are taken below certain thresholds that we now determine.
   
    Let us denote with  $\varphi$ the typical order of magnitude of field values.   Then $\phi_-$ feels a potential of the form $v(\varphi) \equiv m^2\varphi^2/2 - V(\varphi) = m^2\varphi^2/2 -\lambda\varphi^4/4$. This is a potential with a local minimum at $\varphi = 0$ and two maxima at $\varphi = \pm m/\sqrt{\lambda}$ with potential barriers $v (\pm m/\sqrt{\lambda}) = m^4/(4\lambda) \equiv E_d^4$, where we have introduced  
   \be E_d\equiv \frac{m}{(4\lambda)^{1/4}}.\label{Eddef}\ee  
   Since $E_d$ gives the height of the barrier, when the typical energy scale $E$ associated with the field {\it derivatives}  is much below $E_d$ the runaways are avoided. Thus the condition on the  field derivatives to ensure that the motion is bounded is
  \be   \boxed{E\ll E_d}  \qquad \mbox{(condition on field derivatives)}. \label{Condd}\ee
There is a different condition on the field values that can be computed by equating the two terms (the stable and the unstable one) in $v$:  that is $m^2\varphi^2/2 = \lambda\varphi^4/4$, which gives  $\varphi$ equal to 
\be     E_f\equiv \frac{m}{\sqrt{\lambda/2}},\label{Efdef}  \ee
which is larger than $E_d$ for small $\lambda$.
 Thus the condition on the field values to have a bounded motion is 
\be \boxed{\varphi \ll E_f} \qquad \mbox{(condition on field values)}. \label{Condf} \ee 
Both~(\ref{Condd}) and~(\ref{Condf}) have to be satisfied by the boundary conditions in order for the motion to be bounded (that is to avoid the Ostrogradsky instabilities). 
An important point is that when $\lambda$ is small both $E_d$ and $E_f$ become larger than the ghost mass, $m$.  Because of the presence of a potential barrier the solution $\phi_+=\phi_- = 0$ is not unstable,  but {\it metastable}. The geometrical meaning of the thresholds $E_d$ and $E_f$ is illustrated in Fig.~\ref{drawPot}, which shows a  typical  potential with a metastable minimum. This resembles the Higgs potential in the Standard Model for the current central value of the top mass.

\begin{figure}[t]
\begin{center}
 \hspace{0.01cm} \includegraphics[scale=0.342]{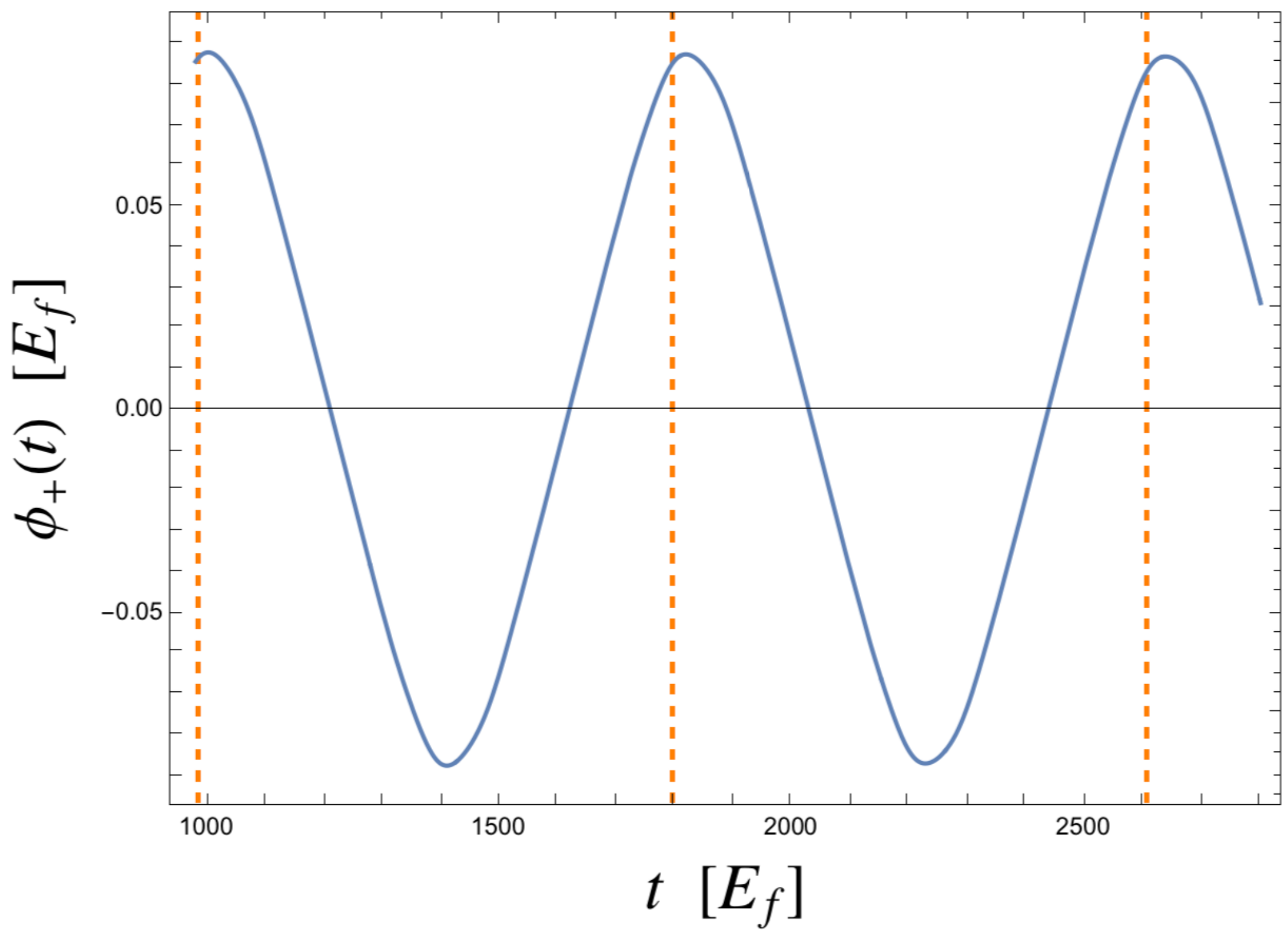}  
  \end{center}
%\label{Sol4derS}
\end{figure}

\begin{figure}[t]
\begin{center}
\includegraphics[scale=0.42]{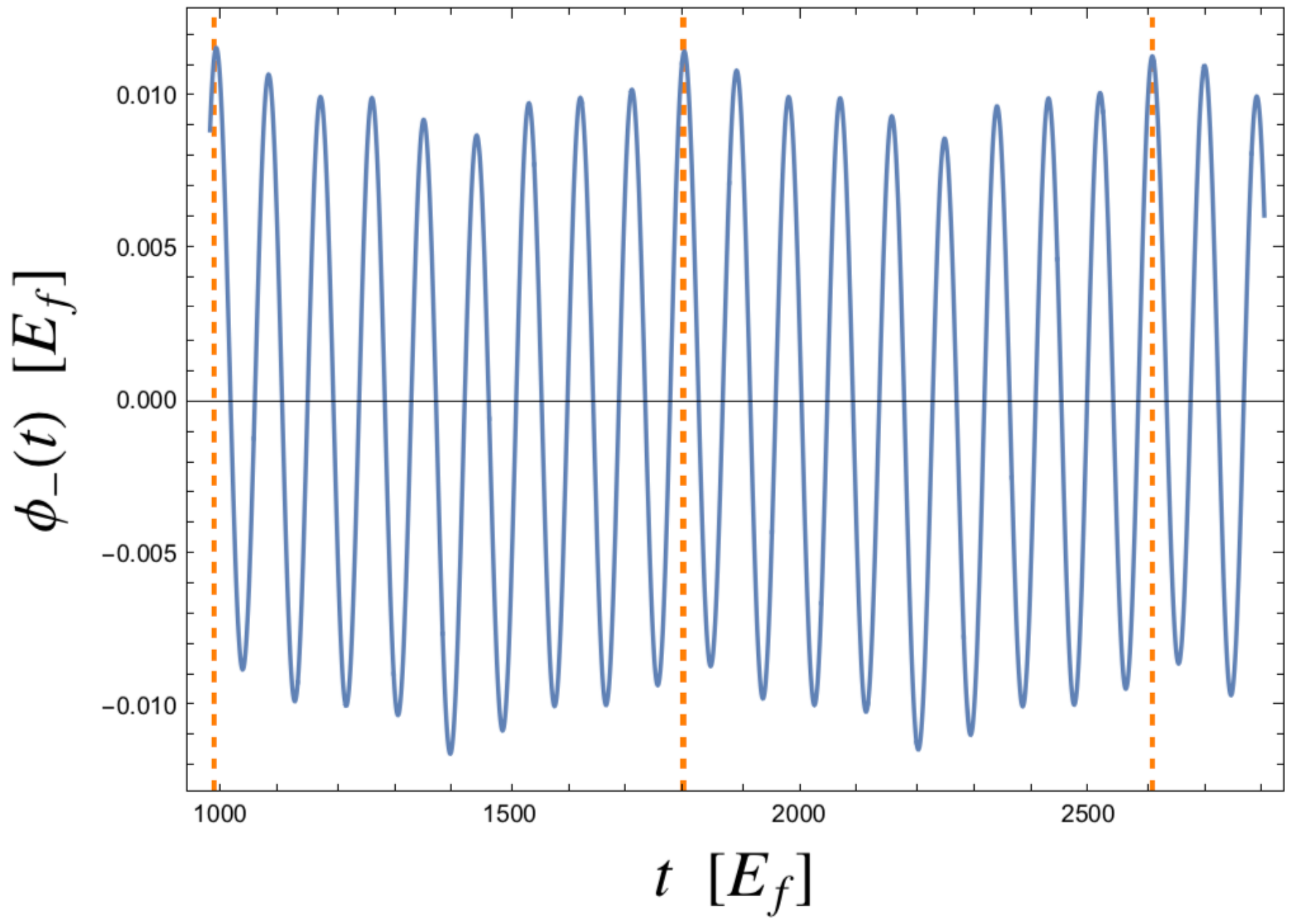} 
  \end{center}
   \caption{\em  Homogeneous time-dependent solution for $V(\phi) = \lambda \phi^4/4$ with $\lambda=10^{-2}$. The initial conditions are chosen as follows: $\phi_+(0) = 10^{-2} E_f$, $\phi_-(0) = 10^{-2} E_f$, $\dot \phi_+(0) = (1.5 \cdot  10^{-1}E_d)^2$ and $\dot \phi_-(0) = -  (10^{-2}E_d)^2$, where a dot denotes the derivative with respect to time $t$. The vertical dashed lines indicate the period.}
\label{Sol4derS}
\end{figure}

In Fig.~\ref{Sol4derS} we show a spatially homogeneous but time-dependent solution for the simple quartic interaction $V(\phi) = \lambda \phi^4/4$. Whenever the initial conditions are chosen to satisfy the bounds in~(\ref{Condd}) and~(\ref{Condf}) we observe indeed a bounded motion (the runaways are avoided) like in the plot\footnote{The existence of an ``island of stability" was noted in the simple 1D toy model obtained from~(\ref{Lag-simple}) by neglecting the spatial derivatives and choosing $V(\phi) = \lambda\phi^4/4$~\cite{ClassToy}. However, Refs.~\cite{ClassToy}  did not identify the thresholds~(\ref{Eddef})-(\ref{Efdef}). 

}. Note that the small values of $\phi_{\pm}$ on the vertical axes just reflect the fact that Condition~(\ref{Condf}) is enforced.

\subsection{The case of quadratic gravity}\label{The case of quadratic gravity}
 
Having illustrated our argument in a simple theory, let us now turn to QG. In analogy with what done in the previous subsection we start by rewriting the $R^2$ and $W^2$  terms as two extra 2-derivative fields.
 
Let us first perform the field redefinition
\be g_{\mu\nu} \rightarrow \frac{\bp^2}{f}g_{\mu\nu} ,\qquad f\equiv \bp^2+\xi_{ab}\phi^a\phi^b -\frac{2R}{3f_0^2}>0,\label{ToGoEinstein}\ee
where the Ricci scalar above is computed in the Jordan frame metric (the one before the redefinition).
Transformation (\ref{ToGoEinstein}) gives the  Einstein frame action~\cite{Kannike:2015apa,Salvio:2018crh,Salvio:2015kka,Salvio:2015jgu}
\be S=\int d^4x\sqrt{-g}\left(- \frac{W^2}{2 f_2^2}-\frac{\bp^2}{2}R  + \mathscr{L}_m^E\right).\label{EinFram}\ee
The Einstein-frame matter Lagrangian, $\mathscr{L}_m^E$, also contains an effective scalar $\omega$, which corresponds to the $R^2$ term in~(\ref{S}) and is defined in terms of $f$ by
\be   \omega =\sqrt{6}\bp \ln\left( \sqrt{f}/\bp\right) .\ee 
 The part of the Lagrangian that depends only on $\omega$ is given by
\be \mathscr{L}_m^\omega = \frac{(\partial\omega)^2}{2} - U, \quad U= \frac{3f_0^2 \bp^4}{8}\left(1-e^{-2\omega/\sqrt{6}\bp}\right)^2. \label{Staro}\ee
The complete form of the Einstein-frame matter Lagrangian, which includes the most general matter sector, can be found in~\cite{Salvio:2018crh} (see also~\cite{Kannike:2015apa}, where the reheating in this class of theories has been studied).

It is also possible to make the ghost explicit by considering an auxiliary field\footnote{We extend the analysis of~\cite{Hindawi:1995an} to a generic matter sector.} $\gamma_{\mu\nu}$:
\bea S&=&\intg \bigg[ 
 \frac{M_2^2\bp^2}{8}\left(\gamma_{\mu\nu}\gamma^{\mu\nu}-\gamma^2\right)\nonumber \\ &&\hspace{1.3cm}-\frac{\bp^2}{2}G_{\mu\nu}\gamma^{\mu\nu} - \frac{\bp^2}{2}R   +\mathscr{L}_m^E  \bigg], \label{two-der}\eea
where $G_{\mu\nu}$ is the Einstein tensor and $\gamma\equiv\gamma_{\mu\nu}g^{\mu\nu}$. Eq.~(\ref{two-der}) can be proved simply by noting that if we insert the solution of the $\gamma_{\mu\nu}$-equations,
\be G_{\mu\nu}=\frac{M_2^2}{2}\left(\gamma_{\mu\nu}-\gamma g_{\mu\nu} \right),\label{pi-eqs}\ee
that is
\be \gamma_{\mu\nu} = \frac{2}{M_2^2}\left(R_{\mu\nu}-\frac{g_{\mu\nu}R}{6} \right), \nonumber  \ee
in (\ref{two-der}) we recover (\ref{EinFram}) (modulo total derivatives). Expanding around the flat metric $\eta_{\mu\nu}$ gives a mixing between $h_{\mu\nu}\equiv g_{\mu\nu}-\eta_{\mu\nu}$ and $\gamma_{\mu\nu}$ that can be removed by expressing $h_{\mu\nu}=\bar h_{\mu\nu} -\gamma_{\mu\nu}$. The tensors $\bar h_{\mu\nu}$ and $\gamma_{\mu\nu}$ represent the graviton and the ghost, respectively. 

Eq.~(\ref{two-der}) is useful because allows us to understand the mass and interaction terms of the ghost. For example, it tells us that the ghost interactions vanish as $f_2\to 0$. This can be seen by inserting~(\ref{pi-eqs}) into~(\ref{two-der}) and by noting that the $f_2\to 0$ limit of the result gives back general relativity (GR). 

Let us consider first the term $\frac{M_2^2\bp^2}{8}\left(\gamma_{\mu\nu}\gamma^{\mu\nu}-\gamma^2\right)$ in~(\ref{two-der}). If one rewrites it in terms of $\bar{h}_{\mu\nu}$ and $\gamma_{\mu\nu}$ it leads  to mass and interaction terms of the schematic form
 \be \frac{M_2^2}{2}\left(\phi_2^2 + \frac{\phi_2^3}{\bp}+  \frac{\phi_2^4}{\bp^2} 
 + ... \right), \ee
 where we have understood  Lorentz indices and order one factors and denoted the spin-2 fields with $\phi_2$ (which we also canonically normalized: $\phi_2\to \phi_2/\bp$).
 The mass term has the same order of magnitude of the interactions for $\phi_2\sim \bp$,  which, therefore, represents the maximal spin-2 field value to avoid the runaways. This maximal value gives $M_2^2 \phi_2^2/2= M_2^4/f_2^2 \equiv E_2^4$, where 
 \be E_2 \equiv \frac{M_2}{\sqrt{f_2}} = \sqrt{\frac{f_2}{2}} \bp. \label{Instability-scale} \ee
{\it For energies}
\be \boxed{E\ll E_2}\quad  \mbox{(in the spin-2 sector)} \label{Cond2}\ee
{\it the Ostrogradsky instabilities are avoided.} Indeed, if~(\ref{Cond2}) is satisfied the mass of the ghost dominates its interactions and, as well-known, a decoupled  ghost does not suffer from runaways.  Condition~(\ref{Cond2}) applies to the {\it derivatives} of the spin-2 fields because, as we have seen above, the threshold for the spin-2 field values is much larger (of order $\bp$). If Condition~(\ref{Cond2}) is satisfied by the boundary conditions the runaways are avoided. The quantity $E_2$ is  analogous to $E_d$ in the simple scalar field theory of Sec.~\ref{A 4-derivative scalar field example}.

   %read up to here
   
The same result holds if one considers the second and third terms in~(\ref{two-der}). The ghost interactions from them have at most the order of magnitude  $f_2 E^2\phi_2^n/\bp^{n-2}$ ($n=1,3,4, ...$),
%n from 2?
 where we have introduced at least a factor of $f_2$ (the ghost decouples for $f_2\to 0$) and the energy squared $E^2$ appears because of the two derivatives present there. These interactions for $\phi_2\lesssim \bp$ are smaller than $\sim f_2 E^2 \bp^2$ that gives $\sim M_2^2\bp^2 \sim (M_2/\sqrt{f_2})^4 = E_2^4$ when evaluated at $E=M_2/\sqrt{f_2}\equiv  E_2$.   
%An alternative way to see this is by using (\ref{pi-eqs}) and noting that the second term 

Finally, let us consider $\mathscr{L}_m^E$. The ghost-matter interactions have size of order 
\be f_2 E^4 \phi_2^n/\bp^n \qquad (n=1,2,3, ...), \label{gm-int}\ee
where now one should interpret $E$ as due to either derivative or mass terms of matter fields or  matter field values times coupling constants; requiring these interactions to be less than $E_2^4$ and  $\phi_2\lesssim \bp$ one obtains that runaways are avoided for
\be \boxed{E\ll E_m,} \quad E_m \equiv \sqrt[4]{f_2}\bp,\quad  \mbox{(in the matter sector)},\label{Condm}\ee 
 which is larger than $E_2$ for small $f_2$. Indeed,~(\ref{Cond2}) regards only the energy $E$ in the spin-2 sector (where the ghost is).  Condition~(\ref{Condm}), like~(\ref{Cond2}), should be satisfied by the boundary conditions. Notice that multiplying $E_m$ by $\sqrt[4]{f_2}$ (to obtain the size of the ghost-matter interactions for the maximal field values, see Eq.~(\ref{gm-int})) one obtains (as one should) a scale of order $E_2$.  The quantity $E_m$ vaguely corresponds to $E_f$ in the simple scalar field theory of Sec.~\ref{A 4-derivative scalar field example}. There is a difference, however: gravity is sourced by other fields (the matter sector) while the scalar $\phi$ of Sec.~\ref{A 4-derivative scalar field example} was assumed to be sourced by itself.\footnote{One could make the two cases more similar by adding  to the theory of Sec.~\ref{A 4-derivative scalar field example} other fields, which mimic the matter sector in QG. We do not do it here for the sake of simplicity.}
 
Therefore, one finds that the ghost of quadratic gravity is not associated with Ostrogradsky instabilities, but rather to metastability: there exist an energy barrier (given in Eqs.~(\ref{Instability-scale}) and~(\ref{Condm})) that prevents the fields from runaway. 
 
 \vspace{-0.cm}
 
 {\it Note that the thresholds $E_2$ and $E_m$ are both larger than $M_2$ for a weakly coupled ghost.} This leads us to a very  interesting situation: there exists an energy range in which the predictions of QG deviates from those of GR, but without activating runaway solutions. We will see some of these predictions~in Sec.~\ref{linear} below. Nevertheless, it is important to observe that Conditions~(\ref{Cond2}) and~(\ref{Condm}) apply to any positive choice of $f_2$. If one takes $f_2\sim 1$ both $E_2$ and $E_m$ are at the Planck scale and again the runaways are avoided for any physical situation that has a chance to be observable. This effect is due to the fact that $M_2$ increases when $f_2$ grows, as clear from~(\ref{M2def}). The disadvantage of the $f_2\sim 1$ case is that, unlike  the $f_2\ll 1$ case, there is no hope to have a large energy window in which there are observational consequences of the ghost without Ostrogradsky instabilities.
 
 For a natural Higgs mass 
 ($f_2\lesssim 10^{-8}$, $M_2 \lesssim 10^{10}$~GeV) 
$E_2$ and $E_m$ can still be as high as $10^{-4}\bp$ and  $10^{-2}\bp$, respectively.
 {\it It is clear that inflation (and the preceding epoch) is the only stage of the universe that can provide us information about such high scales.}  For energies much below $M_2$ (which can be many orders of magnitude above the TeV scale for a natural Higgs mass and is at the Planck scale for $f_2 \sim 1$) the theory reduces to Einstein gravity and, therefore, all the observations related to low energy astrophysical systems (involving typical energies much smaller than $M_2$) are reproduced just as in Einstein gravity.  Let us then focus on inflation and the pre-inflationary epoch.
 
 Note that  $W_{\mu\nu\rho\sigma}=0$  on a Friedmann-Robertson-Walker (FRW)
metric because such metric is conformally flat. Therefore, only perturbations that violate homogeneity and/or isotropy could destabilize the universe. In the chaotic  theory~\cite{Linde:1983gd} (a key element to understand the naturalness of inflation) one assumes that the fields took random values including inhomogeneous and anisotropic ones before inflation. But we live in one of those patches where the energy scales of inhomogeneities ($1/L$) and anisotropies ($A$) were small enough~\cite{Weinberg:2008zzc}:
\be  L \gg |\Phi/U'(\Phi)|^{1/2}, \qquad A\ll H, \label{ChaosCond}\ee
where $H$ is the inflationary Hubble rate, $\Phi$ is a generic canonically normalized inflaton field and $U$ is its potential. The conditions above justify the use of homogeneous and isotropic solutions to describe the classical part of inflation (and is  regularly done in the literature on inflation). On the one hand, the experimental bound~\cite{Ade:2015lrj} 
\be H< 2.7\cdot 10^{-5}\bp \quad  (95\%~ \mbox{CL})\label{ObsBound} \ee
implies that the second condition in~(\ref{ChaosCond}) ensures both~(\ref{Cond2}) and~(\ref{Condm})  (at least for the maximal ghost mass compatible with Higgs naturalness). On the other hand, by identifying $U$ with the one in~(\ref{Staro}) (Starobinsky inflation\footnote{We find similar results with other successful models such as Higgs inflation~\cite{Bezrukov:2014bra} or Hilltop inflation~\cite{Boubekeur:2005zm}.}~\cite{Starobinsky:1980te}) the first condition in~(\ref{ChaosCond}) becomes $1/L\ll 10^{-6}\bp$, which again agrees with  both~(\ref{Cond2}) and~(\ref{Condm}).

 {\it In other words, remarkably, the chaotic theory automatically ensures that the conditions to avoid runaway solutions are satisfied.}

  \section{Explicit nonlinear calculations}\label{nonlinear}

In this section we solve the nonlinear gravity equations,\footnote{From now on we perform the calculations in the Einstein frame~(\ref{EinFram}) unless otherwise stated. }  namely
  \be G_{\mu\nu}+\frac2{M_2^2} B_{\mu\nu} = \frac{T^E_{\mu\nu}}{\bp^2},  \ee
where  $B_{\mu\nu}\equiv  \left(\nabla^\rho\nabla^\sigma +\frac{R^{\rho\sigma}}{2}\right) W_{\mu\rho\nu\sigma}$ is the Bach tensor and $T^E_{\mu\nu}$ is the energy-momentum tensor in the Einstein frame. We do so with an ansatz that violates the symmetries of the FRW metric to see the nonlinear effect of the ghost.
  In order to understand how things work in practice, we consider the following  anisotropic ansatz\footnote{For the study of other anisotropic metrics in QG and conformal gravity see Refs.~\cite{Cotsakis:1993ezo,Barrow:2005qv} and~\cite{Demaret:1998dm}, respectively.},
  \be ds^2 = dt^2 -a(t)^2 \sum_{i=1}^3 e^{2\alpha_i(t)}dx^i dx^i.  \label{anisotropicmetric} \ee
The scale factor $a$ and the $\alpha_i$ are generic functions of the cosmic time, $t$. As usual $H$ is defined in terms of $a$ by $H\equiv \dot a/a$, where a dot denotes the derivative with respect to $t$. This allows us to probe the nonlinear dynamics by solving ordinary differential equations.   Note, however, that the argument of the previous section also applies to inhomogeneities. To the best of our knowledge this is the first article where anisotropic metrics in QG are studied in the Einstein frame, which allows us to compare theoretical predictions with inflationary observations (as we will do in Sec.~\ref{linear}). In this section we focus on Starobinsky inflation,  a natural option in QG.

Given that $\sum_i \alpha_i$ can be included in a redefinition of $a$ we can take
\be \alpha_1 \equiv \beta_+ + \sqrt{3}\beta_-, \quad \alpha_2 \equiv \beta_+ - \sqrt{3}\beta_-, \quad \alpha_3 = -2 \beta_+. \ee
Therefore, the amount of anisotropy is encoded only in  the functions $\beta_{\pm}$  and we can measure it through
\be A\equiv \sqrt{\dot\beta^2_++\dot\beta^2_-}.  \label{anisotropyDef}  \ee

By inserting this ansatz in the trace of the gravity equations one obtains
\be R= \frac{3p -\rho}{\bp^2},\qquad  \bigg(\rho = \frac{\dot\omega^2}{2} +U, \quad   p = \frac{\dot\omega^2}{2} -U\bigg). \label{Rfixed}\ee
The Weyl-squared term does not contribute to this equation.
If the anisotropy is zero,  the other equations do not receive contribution  either. However, for $A\neq 0$ the Weyl-squared term contributes to some equations and leads to terms with 4 derivatives. One can reduce the gravity system   to first-order equations through the definitions 
\be \gamma_\pm = \dot\beta_\pm, \quad \delta_\pm = \dot\gamma_\pm, \quad   \epsilon_\pm =\dot\delta_\pm\label{gde}.\ee 
The $tt$-component of the gravity equations then is
\bea H^2&=&\frac{\rho}{3\bp^2}+A^2-\frac1{M_2^2} \left[ \frac{R A^2}{3}+H^2 A^2+  \right.\nonumber \\
&&\hspace{-2cm} \left.14 A^4-4H (\gamma_+ \delta_++\gamma_-\delta_-)  -2 (\gamma_+ \epsilon_++\gamma_-\epsilon_-) + \delta_+^2 + \delta_-^2\right]\hspace{-0.05cm}.  \label{Heq} \eea
Note that, by using~(\ref{Rfixed}), this equation becomes a second-order algebraic equation for $H$. As usual we choose the solution that supports the universe expansion.  The $ii$-components of the gravity equations lead instead to the $\epsilon_\pm$-equations: 
\bea  \dot \epsilon_\pm  &= & - M_2^2\left(3 H \gamma_\pm+\delta_\pm \right)  \nonumber \\
&&+\left[\frac{\dot R}{6}+\frac{RH}{2}+27HA^2+18(\gamma_+\delta_++\gamma_-\delta_-)\right] \gamma_\pm \nonumber \\ && +\left(\frac{2R}{3}-3H^2+12 A^2\right) \delta_\pm -6H\epsilon_\pm, \label{epsilonEq}\eea
where we have used the $tt$-component and the trace of the gravity equations.
% to eliminate $\rho$ and $p$. 
Finally, we also reduce the inflaton equation to two first-order differential equations:
\be  \dot\omega =  \pi_\omega, \qquad \dot \pi_\omega + 3H\pi_\omega = -\frac{dU}{d\omega}. \label{omegaEq}\ee
By using the first one of these equations and Eq.~(\ref{Rfixed}) one can express $R$ in terms of $\omega$ and $\pi_\omega$.
Eqs.~(\ref{gde})-(\ref{omegaEq}) then form a set of 11 equations in 11 unknowns ($\beta_\pm$, $\gamma_\pm$, $\delta_\pm$, $\epsilon_\pm$, $H$, $\omega$, $\pi_{\omega}$). We find, as we should, that there are no other independent equations.

\begin{figure}[t]
\begin{center}
  \includegraphics[scale=0.24]{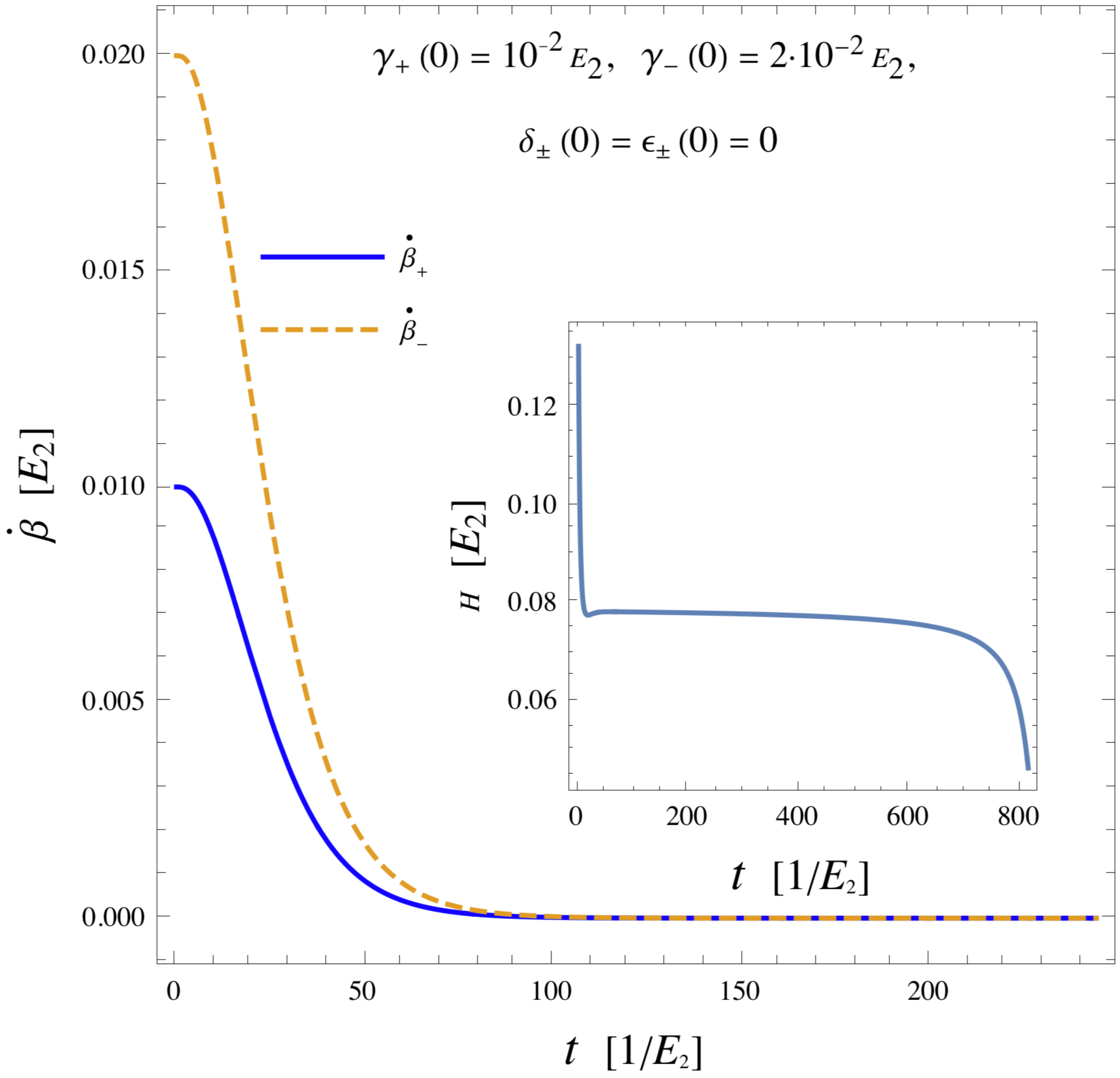}
  %5
  \end{center}
   \caption{\em The anisotropy versus the cosmic time. We set $f_2=10^{-8}$, $f_0 \approx 1.6 \cdot 10^{-5}$, $\phi(0)\approx 5.5 \bp$ and $\sqrt{\pi_\phi(0)}\approx 7.1 \cdot 10^{-6}\bp$. In the inset we show the corresponding  Hubble rate.}
\label{AniTime}
\end{figure}

In Fig.~\ref{AniTime} we show how small initial values for the anisotropy in the sense of~(\ref{Cond2})-(\ref{Condm})  ($|\gamma_\pm(0)|\ll E_2$, $\sqrt{|\delta_\pm(0)|}\ll E_2$, $\sqrt[3]{|\epsilon_\pm(0)|}\ll E_2$ and\footnote{By using the definitions in~(\ref{Instability-scale}) and~(\ref{Condm}) one can equivalently write the last condition $H\ll E_2$ as $\sqrt{\bp H}\ll E_m$. This is because we can also regard $\bp H$ as generated (through the Einstein equations) by the energy density stored in the matter sector.} $H\ll E_2$) do not create problems: the anisotropy quickly goes to zero and one recovers the GR behavior; no runaway solutions are observed in agreement with the general argument of Sec.~\ref{model}.  Note that the smallness of $\dot\beta_{\pm}$ simply reflects the fact that Condition~(\ref{Cond2}) is enforced because the plot is presented in units of $E_2$; for the chosen value of $f_2$, the quantities $\dot\beta_{\pm}$ in units of $M_2$ are actually much larger than one at the beginning. Regarding $H$, after a short (preinflationary) time, it quickly reaches a plateau (inflation) and then decreases again when inflation ends.  The inflaton initial conditions in  Fig.~\ref{AniTime} were chosen to obtain $N \approx 62$. We scanned the possible values of initial conditions, not just the one used in Fig.~\ref{AniTime}, and always found qualitatively the same  result. 

\begin{figure}[t]
\begin{center}
  \includegraphics[scale=0.25]{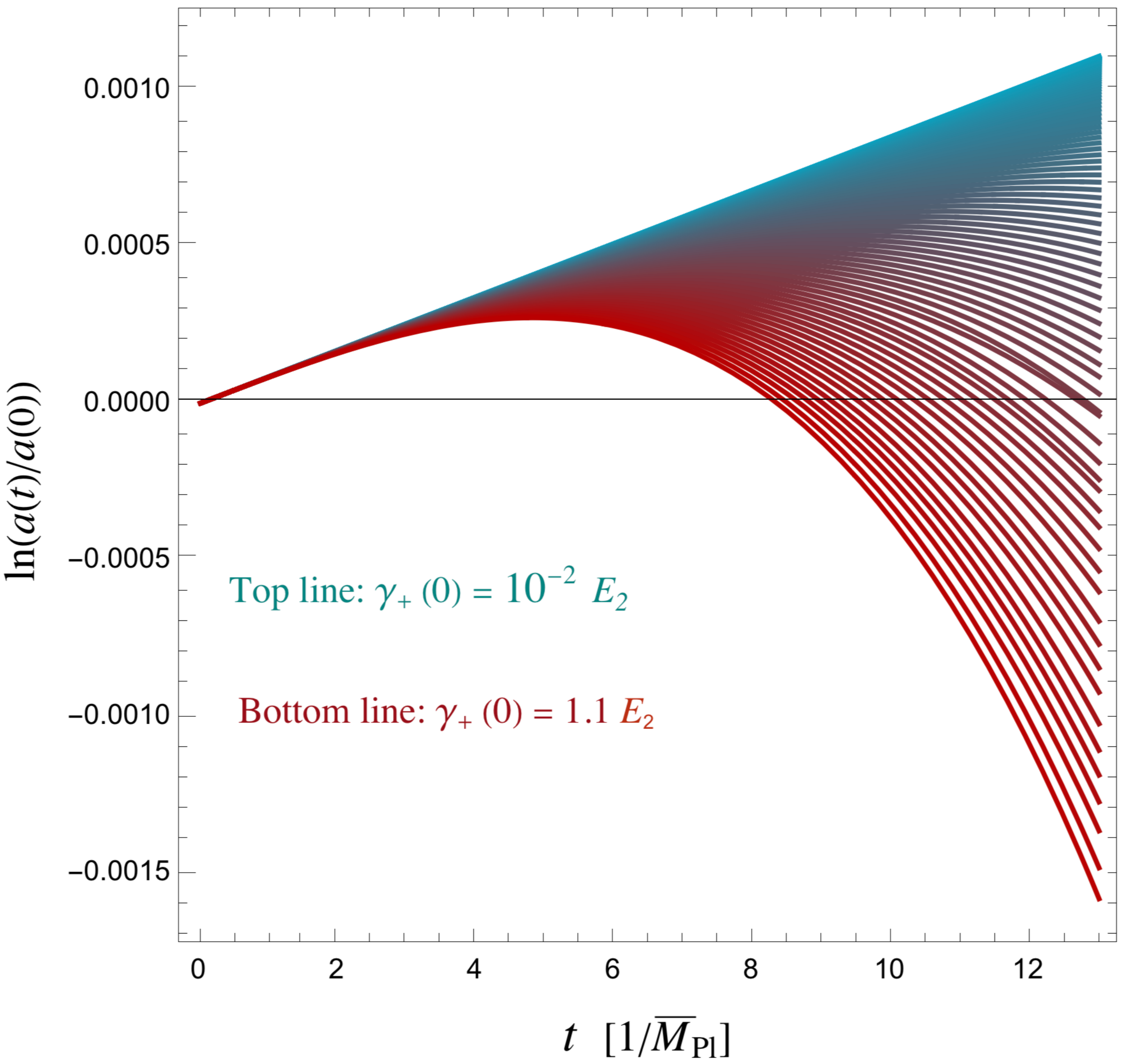}
  %52
  \end{center}
   \caption{\em The scale factor in the Jordan frame by varying the initial condition for $\gamma_+$. We set $\gamma_-(0)= 10^{-1}E_2$ and all the other initial conditions for the anisotropy functions to zero. Furthermore, $f_2=10^{-8}$, $f_0 \approx 1.6 \cdot 10^{-5}$, $R(0) \approx 1.3 \cdot 10^2 f_0^2 \bp^2$ and $H(0) = 1.2E_2$.}
\label{InfColl}
\end{figure}

For the considered anisotropies we find that when~(\ref{Cond2})-(\ref{Condm})  are {\it not} satisfied the universe collapses as shown in\footnote{The scale factor in Fig.~\ref{InfColl} is the one in the Jordan frame because we find that the field redefinition in~(\ref{ToGoEinstein}) is ill-defined when collapse occurs (the inequality in~(\ref{ToGoEinstein}) is not satisfied for some time). } Fig.~\ref{InfColl}. In that plot we have set all the initial conditions  of the anisotropy functions (other than $\gamma_{\pm}(0)$) to zero; by turning on  the other initial conditions one finds similar behaviors: when the energy scales associated with them are much smaller than $E_2$ the anisotropy goes to zero as time passes by and the universe inflates; when they are comparable or larger than $E_2$ the collapse occurs. 

The interpretation is that the regions of space with large  initial  anisotropies have eventually zero size as compared to those satisfying~(\ref{Cond2})-(\ref{Condm}), which instead lead to inflation. What we have shown provides an explicit mechanism to implement the original chaotic inflation idea by Linde~\cite{Linde:1983gd}. Indeed, it is not  clear if in Einstein gravity the patches that were largely inhomogeneous and anisotropic (where Conditions~(\ref{ChaosCond}) were violated) are incompatible with life: the fact that inflation does not occur in that case is not sufficient to reach this conclusion. On the other hand, the classical runaways that are triggered when those conditions are violated in the presence of the $W^2$ term for a natural Higgs mass ($f_2\sim 10^{-8}$) certainly render   the universe inhospitable. Indeed, it is very interesting to note that the maximal Hubble rate allowed by~(\ref{ObsBound}) is, remarkably,  just slightly smaller than the value of  $E_2$ for the natural choice, $f_2\sim 10^{-8}$. Therefore, the Weyl-squared term combined with Higgs naturalness provides an explicit mechanism to implement the original chaotic inflation idea.

\section{Linear analysis and observational predictions}\label{linear}

A general check of ghost metastability can be performed by studying the complete set of linear perturbations around the de Sitter (dS) spacetime (which is the relevant one according to the results of Sec.~\ref{model}): one should find no runaway solutions there for arbitrary energies. The linear dS modes were found in~\cite{Salvio:2017xul,Ivanov:2016hcm,Clunan:2009er,Deruelle:2010kf,Deruelle:2012xv,Salles:2014rua,Myung:2014jha,Myung:2014cra,Shapiro:2014fsa,Myung:2015vya,Salles:2017xsr}. Here we show that they are  all bounded (and thus they do not suffer from runaways) for any wave number  $q$.

First, one should recall that for $M_2>H$ one recovers the Einsten modes, which are bounded. This expected decoupling was rigorously shown in Ref.~\cite{Salvio:2017xul}.
 Therefore, we focus here on the case $M_2<H$, which, taking into account~(\ref{ObsBound}), implies  that $f_2$ has to be very small. In the following we choose to work in the conformal Newtonian gauge.

Let us start with the scalar perturbations. They are like in GR with one exception: there is one more isocurvature mode $B$ (the helicity-0 component of the ghost)~\cite{Salvio:2017xul,Ivanov:2016hcm}. Its modes are $g_B$ and $g_B^*$, where\footnote{In~\cite{Ivanov:2016hcm} it was proved that no physical singularity  can be present in any other gauge. This proof was later extended to the most general matter sector in Ref.~\cite{Salvio:2017xul}.}~\cite{Salvio:2017xul}
\be   g_B(\eta, q) \equiv \frac{H}{\sqrt{2q}} \left(\frac{3}{q^2}+\frac{3i\eta}{q}-\eta^2\right)e^{-iq\eta} + {\cal R}\mbox{-terms} \label{gBmodes}\ee
and $\eta$ is the conformal time ($\eta<0$ and $\eta\to 0^-$ corresponds to large $t$). 
%$a^2d\eta^2=dt^2$, 
The terms due to the curvature perturbation ${\cal R}$ are not shown since they are the same as in GR and thus are bounded. Also the first term in~(\ref{gBmodes}) is bounded: for given initial conditions at a negative value of $\eta$ the superhorizon limit $\eta\to 0$  is finite.

\begin{figure}[t]
\begin{center}
  \includegraphics[scale=0.16]{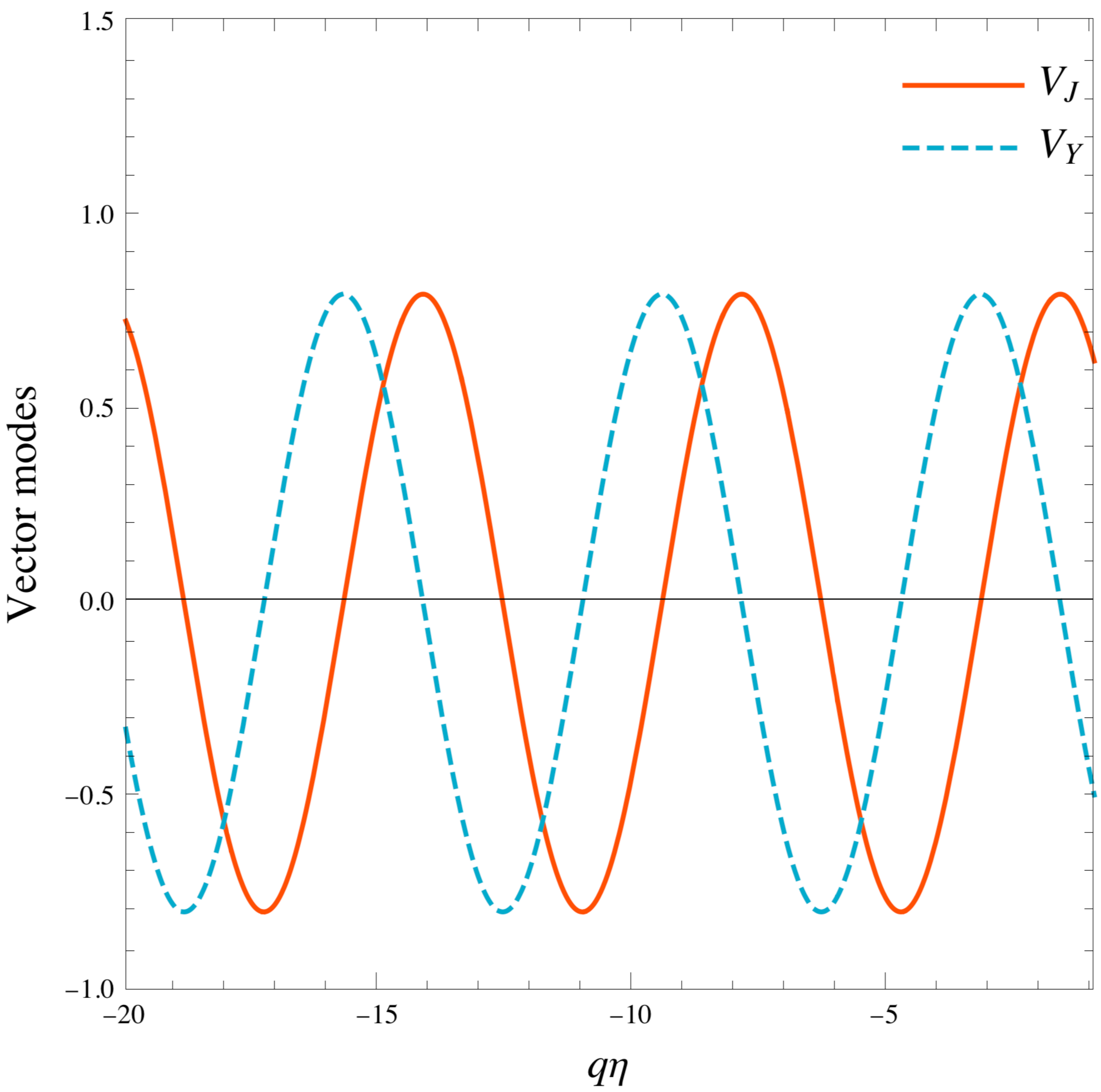}
  \vspace{-0.4cm}
  
  %4
     \end{center}
   \caption{\em Vector modes for $\rho\equiv H^2/M_2^2 = 10^4$. }
\label{Vecf1}
\end{figure}

The vector modes are instead given by\footnote{For the derivation of the vector modes see~\cite{Salvio:2017xul} , where a previous calculation in~\cite{Myung:2014jha} was corrected.} 
\be  V_I\equiv   \sqrt{-q\eta}\,\,  I_{\frac{\sqrt{\rho -4}}{2 \sqrt{\rho }}}(-q \eta), 
\ee
where $I=\{J, Y\}$, $J_\alpha$ and $Y_\alpha$ are the Bessel functions of the first and second kind, respectively, and $\rho\equiv H^2/M_2^2$. We plot them in Fig.~\ref{Vecf1} to show they are bounded.

\begin{figure}[t]
\begin{center}
  \vspace{0.2cm}
  \includegraphics[scale=0.19]{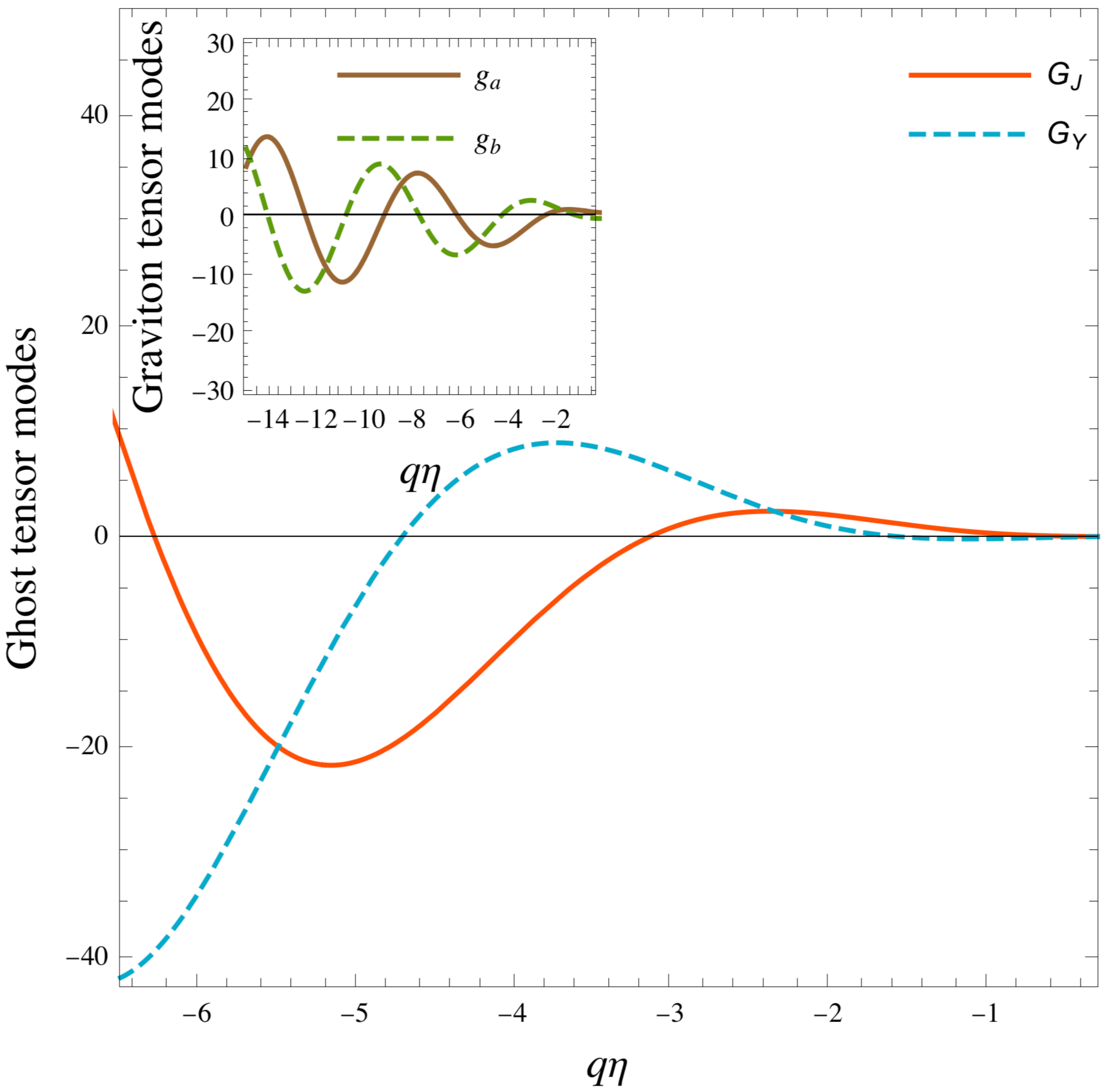} 
  %45
    \vspace{-0.4cm}
  \end{center}
   \caption{\em Tensor modes for $\rho\equiv H^2/M_2^2 = 10^4$. 
   }
\label{VecT1}
\end{figure}

 In the tensor sector we have four modes (the  two helicity components of the  graviton and the ghost). They are~\cite{Clunan:2009er,Deruelle:2012xv,Salvio:2017xul}
\be g_a \equiv  \cos (q\eta) +q\eta \sin (q\eta), \quad g_b \equiv  q\eta \cos (q\eta)-\sin (q\eta), \ee
\be G_I \equiv (-q\eta)^{3/2} I_{\frac{\sqrt{\rho -4}}{2 \sqrt{\rho }}}(-q \eta),  \quad 
 (\mbox{where}\,\,  I=\{J, Y\})
 \ee
 for the graviton ($g$) and the ghost ($G$) respectively. The graviton tensor modes $g_a$ and $g_b$ coincide with the modes one finds in GR. In Fig.~\ref{VecT1} we show that the ghost tensor modes $G_Y$ and $G_J$ are also bounded (we plot the corresponding graviton tensor modes too for comparison).

In Figs.~\ref{Vecf1} and~\ref{VecT1} we chose $\rho=10^4$, which is a typical value in QG with a natural Higgs mass. The results do not change qualitatively as long as $H\gg M_2$. For ghost masses above $H$ one instead recovers the modes of GR.

 Note that the modes presented in this section would reduce to a {\it linear} analysis of the ansatz in (\ref{anisotropicmetric}) if one considers the space-independent limit of the perturbations  and  matches the two different definitions of time used here and in Sec.~\ref{nonlinear}.

%[important:]
%[For f2 very small (necessary in order not to fall in the Einstein case) it has been shown in the perturbation paper that \Phi =-\Psi. For space independent perturbations this gives an extra contribution to the relation between t and eta. So in the linear limit the Bianchi 1 setup can be recovered by considering the space independent limit of the perturbations and expressing the conformal time in terms of the cosmic time taking into account this extra contribution.]
%

The linear analysis also provides observational predictions of the theory. There are two differences compared to GR as shown in~\cite{Salvio:2017xul}. The first one is a suppression of the tensor power spectrum  such that the tensor-to-scalar ratio $r$ is\footnote{The power spectra are at horizon exit $q=a H$.}
$r=r_E/(1+2 H^2/M_2^2)$,
where $r_E$ is the tensor-to-scalar ratio in GR. In QG a natural Higgs mass corresponds typically~\cite{LowScale} to  $H\gg M_2$ so $r$ is highly suppressed.
For example, for the parameter values and inflationary model chosen in Fig.~\ref{AniTime} $r_E\approx 0.003$, $r\sim 10^{-9}$,   the curvature power spectrum $P_{\cal R} \approx 2.1 \cdot 10^{-9}$ and $n_s\approx 0.968$. All  predictions are in agreement with the most recent Planck data~\cite{Ade:2015lrj}. 

 \begin{figure}[t]
\begin{center}
\qquad  \includegraphics[scale=0.25]{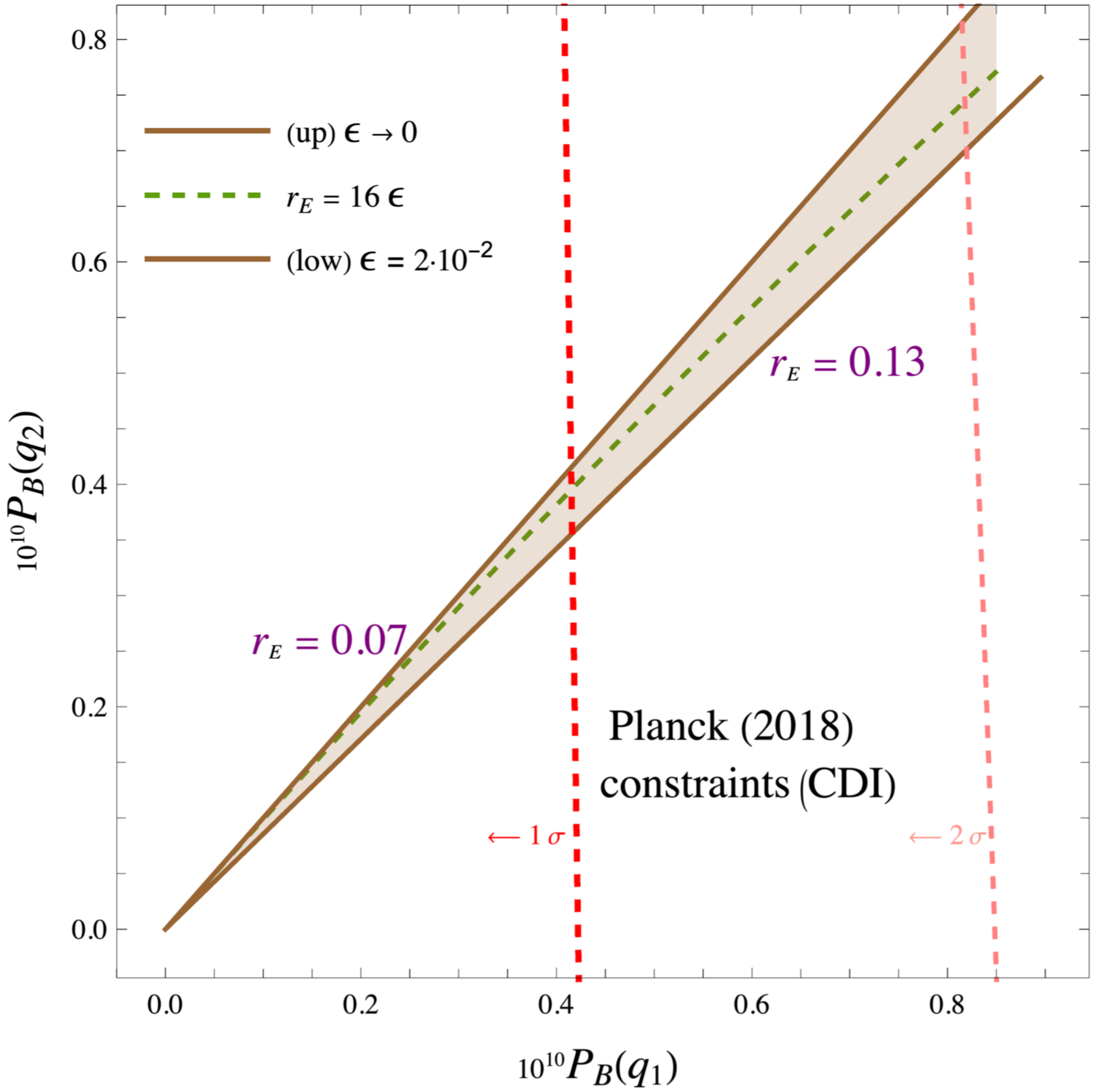} 
  \end{center}
   \caption{\em  The ghost-isocurvature power spectrum $P_B$ computed at two different scales ($q_1 = 0.002~{\rm Mpc}^{-1}$ and $q_2 = 0.1~{\rm Mpc}^{-1}$). The most precise determination of $P_\mathcal{R}(q_0)$ (the curvature power spectrum at the pivot scale $q_0 = 0.05~{\rm Mpc}^{-1}$) by Planck (2018) is used. The strongest constraints from Planck at 1-2$\sigma$ level are also shown.}
\label{PB1vs2}
\end{figure}

 The second difference is the presence of $B$, whose power spectrum is
\be P_B =\frac{3}{2\bp^2} \left(\frac{H}{2\pi}\right)^2.  \label{PBspectrum} \ee 
 Given that $P_B$ is {\it not} suppressed (and in fact it is 3/16 times the tensor power spectrum in GR) we compute here its dependence on $q$, which is required to compare it with the Planck constraints on isocurvature modes~\cite{Ade:2015lrj}. By defining the spectral index as  $n_B\equiv 1 +\frac{d\ln P_B}{d\ln q}$ one finds $n_B = 1-2\epsilon$, where $\epsilon$ is the first slow-roll parameter (in single field inflation $\epsilon=r_E/16$). Then the $q$-dependence  is 
\be P_B(q) = P_B(q_0) \left(\frac{q_0}{q}\right)^{2\epsilon}. \ee
In Fig.~\ref{PB1vs2} we compare $P_B$ with Planck data. As shown in the plot, models with $r_E \approx 0.2$ are compatible with the data, unlike in GR. 
Moreover, given that the spectral index $n_B$ is close to 1, $B$ fulfills the bounds on the spectral index of isocurvature modes given in~\cite{Ade:2015lrj}. The CMB-S4 collaboration will be able to improve the sensitivity to isocurvature modes~\cite{Abazajian:2016yjj} and, therefore, this scenario can be further tested in the future.

\vspace{-0.3cm}

\section{Conclusions}\label{Conclusions}

\vspace{-0.2cm}

We have shown  that the possible classical runaways of quadratic gravity do not occur if the energies satisfy Conditions~(\ref{Cond2}) and~(\ref{Condm}), which regard the boundary conditions for the spin-2 sector  and the matter sector, respectively. Those conditions are weak enough to accommodate the entire history of the universe. For a  natural Higgs mass with $f_2\sim 10^{-8}$, $E_2$ and $E_m$ are still so high that can be tested only via inflation (and pre-inflation dynamics). In that context, those energies represent the deviations from a homogeneous and isotropic metric (given that the ghost is inactive in a conformally flat metric). 

To illustrate how this argument works we have solved numerically the nonlinear equations for anisotropic metrics. We found that the regions where Conditions~(\ref{Cond2}) and~(\ref{Condm}) are satisfied quickly become isotropic and inflate; the others undergo collapse. The possible lethal instabilities occurring whenever the energy bounds are violated not only are avoided in our universe, but would also explain (for a natural Higgs mass) why we live in a homogeneous and isotropic universe: life can only emerge from those patches that are enough isotropic and homogeneous in the sense of~(\ref{Cond2}) and~(\ref{Condm}).

As a check of the general argument, we have also shown that the linear perturbations around dS are bounded for any $q$. Those linear modes also encode important and testable predictions of the theory: most notably a gravity-isocurvature mode that satisfies all current bounds and can be tested with CMB observations in the next future. 

It is also appropriate to mention here some advantages of our approach with respect to an alternative proposed in Refs.~\cite{Anselmi:2018bra,Anselmi:2019rxg} where the ghost is projected out from the spectrum and the classical limit is taken: first our argument holds for generic metrics while the method of~\cite{Anselmi:2018bra,Anselmi:2019rxg} is developed (so far) only for flat and purely FRW metrics (and the general perturbations around FRW are crucial to make contact with observations); second the classical runaways, which appear when  Conditions~(\ref{Cond2}) and~(\ref{Condm}) are violated, gives us an explanation of the quasi-homogeneity and isotropy of the preinflationary initial conditions.

Finally, our results render ligitimate and 
 motivate the study of classical solutions in QG, such as black holes and horizonless spherical solutions (partially explored in~\cite{Lu:2015cqa,Lu:2015psa,Lu:2017kzi,Svarc:2018coe,Goldstein:2017rxn}  and~\cite{Holdom:2002xy,Holdom:2016nek}), wormholes (see~\cite{Hohmann:2018shl} for the conformal gravity case) and gravitational waves.

\vspace{-0.2cm}

\section*{Acknowledgments} 
\vspace{-0.1cm} 

\noindent   I thank B.~Holdom, M.~Ivanov, J.~Ren, S.~Sibiryakov,  M.~Simonovic, A.~Strumia, H.~Veerm\"{a}e and C.~Wetterich
  for useful discussions. I also thank CERN for the hospitality during the last stage of the work.

\vspace{0.9cm}

%\newpage

\end{document}